\documentclass[12pt]{article}
\usepackage{amsmath,amssymb,amsthm,amssymb}
\title{Classical and quantum randomness and the financial market}
\author{Andrei Khrennikov\\
School of Mathematics and Systems Engineering\\
University of V\"axj\"o, S-35195, Sweden}

\begin{document}

\maketitle

\begin{abstract} We analyze complexity of financial  
(and general economic) processes by comparing classical and quantum-like 
models for randomness. Our analysis implies that it might be that a 
quantum-like probabilistic description is more natural for financial market than the 
classical one. A part of our analysis is devoted to study the possibility of
 application of the quantum probabilistic model to agents of financial market. 
We show that, although the direct quantum (physical) reduction (based on using 
the scales of quantum mechanics) is meaningless, one may apply so called quantum-like models. 
In our approach quantum-like probabilistic behaviour is a consequence of contextualy 
of statistical data in finances (and economics in general). However, our hypothesis on 
"quantumness" of financial data should be tested experimentally (as opposed to the 
conventional description based on the noncontextual classical probabilistic approach). 
We present a new statistical test based on a generalization of the well known in quantum 
physics Bell's inequality.
\end{abstract}

\bigskip


\section{Introduction}
The financial market is a complex dynamical system and, 
since the publication of the thesis of L. Bachelier [1], there were performed  
numerous studies devoted to various aspects of random description of financial processes [2]. 
At the first stage of investigations {\it Brownian motion was used to describe randomness of the
financial market}. This model provided a rather good approximation of some financial processes.
However, later it became evident that the diversity of financial stochastic processes could
not be reduced to Brownian motion. The next step was consideration of functionals of Brownian 
motion, especially, geometric Brownian 
motion [2]. Later there were considered other types of stochastic processes [2], 
in particular, general {\it Levy processes}.

But even the possibility {\it to describe financial market by the classical probabilistic model}
 (and, in particular, dynamics of prices by stochastic differential equations) can be questioned 
[3]-[13]. We emphasize that the financial market is not a kind of a mechanical system. It is 
not a direct analog of a huge system of interacting physical particles. \footnote{ Of course,
 one could proceed quite deeply by using such an analogy with statistical mechanics.} 
Expectations of traders, exchange of information, news, political and social events are 
not less important than ``the real development of industry and trade.'' 
It might be more natural to consider financial market as a huge {\it information system} 
(and not mechanical system) [3], [4], [11]--[13]. 
Moreover, we can even speculate that an adequate description of financial market
 can be approached through considering it as a {\it cognitive system} [3], [4], [11]--[13]. 
Hence, it may be possible to use the experience of mathematical description of 
cognitive systems for the financial market.

We recall basic approaches to description of brain's functioning. 
There is a very strong tendency to interpret brain's work as functioning of a 
huge {\it neural network}. Such a description (which is based on the classical 
probability theory and its application to the financial market -- the network approach) 
generates classical financial processes. However, some groups of researchers 
in cognitive science and psychology do not believe that brain's work could be 
totally reduced to the exchange of electric impulses between neurons. 
For example, it seems to be impossible to embed the consciousness into the reductionist picture. 
There is no any idea how a neuronal network (even extremely complex) would be able to produce 
the consciousness. It seems that brain's complexity could not be reduced to complexity 
of networks. 

We remark that there were numerous attempts 
to apply quantum mechanics to describe mental processes, see [4] 
for extended bibliography. The main problem of the quantum approach to the description 
of mental processes is the impossibility to combine the neuronal and quantum models. 
The conventional interpretation of quantum mechanics is based on the notion of superposition 
of states for an individual quantum system, e.g.,  an electron. An electron can have a
state of superposition to be in two different places simultaneously. On the
other hand, a neuron could not be at the same time in the states firing and non-firing.

The conventional quantum interpretation of superposition induces
a rather special viewpoint on randomness - {\it individual randomness} 
(see J. von Neumann [14], see also [15], [16]). It is commonly assumed 
that quantum randomness (described by the complex wave function $\psi(x)$) could 
not be reduced to classical ensemble randomness. 
The latter is induced by a variety of properties of elements of a statistical ensemble.
It is described by the classical measure-theoretical approach based on the axiomatics of Kolmogorov.

We remark that A. Einstein, E. Schr\"odinger, L. De Broglie, D. Bohm strongly criticized such a viewpoint to quantum randomness. They were sure that quantum randomness could be reduced to classical ensemble randomness. The main problem in combining quantum probability with classical ensemble probability is to find a reasonable explanation of the interference of probabilities. Instead of the ordinary addition of probabilities of alternatives:
\begin{equation}
\label{1}
P=P_1 + P_2,
\end{equation}
the quantum probabilistic calculus (based on transformations of vectors in the complex Hilbert space) predicts the general rule:
\begin{equation}
\label{2}
P=P_1 + P_2 + 2\cos \theta \sqrt{P_1 P_2}.
\end{equation}
This is the 
so called {\it quantum interference of probabilities}. The difference between the rules (\ref {1}) and (\ref{2}) was considered as the strongest argument against the classical ensemble randomness and in favor of {\it individual quantum randomness}. Therefore a rather common opinion in quantum community is that quantum randomness differs cruicially from the classical one and, in particular, it could not be considered as ensemble randomness.

Such a viewpoint on randomness induces huge difficulties in applications of the 
quanum formalism outside the quantum domain. In particular, in cognitive sciences 
people should go to the quantum scales of space and time (R. Penrose even tried to go to
the scales of quantum gravity). Such attempts of quantum reductionism did not find so
 much understanding by neurophysiologists, psychologists, cognitive scientists who 
did not believe that cognitive phenomena could be explained only at the quantum scale.

It is even more difficult to apply the conventional quantum approach to 
randomness for the description of the financial market and general economic processes. 
In principle, one might believe that the brain functioning could be reduced to 
processes in the microworld (interactions of quantum particles in the brain), 
but it would be a funny thing to try to reduce the functioning of the financial market 
to interactions of photons and electrons, protons and neutrons composing traders of the market.

Recently a solution of the interference problem in the classical ensemble framework was 
proposed in a series of authors papers, see [4] for the general presentation. 
The crucial point is that all (classical) probabilities should be considered as {\it contextual probabilities}. Here by context we understood a complex of conditions: physical, biological, economic or social. By taking into account the dependence of probabilities on contexts we can reproduce all distinguishing features of the  quantum formalism: interference of probabilities, Born's rule (i.e., the possibility to represent probability as the square of the absolute value of a complex amplitude - wave function), the representation of random variables by noncommutative operators. It seems that the essence of quantum formalism is not individual quantum randomness, but contextuality of probabilities.

Such a viewpoint to the quantum probabilistic calculus - the 
quantum-like approach - provides new possibilities for its application. 
There is no more need for looking in economy for mystical objects 
being in superposition of their states. Any context dependent system, 
for instance, agents of market, may exhibit the quantum probabilistic behaviour and, 
hence, should be (in such a case) described by the quantum probability model.

The main difference between the quantum probabilistic description and 
the classical one is that the  dynamical equation is not a stochastic differential equation, 
but the Schr\"odinger equation for the complex probability amplitude - the 
wave function\footnote{ In our approach the wave function is just the special  probabilistic 
representation of a context.}. Such a quantum-like approach provides an approximative probabilistic representation of some class of processess in the situation in which the complete probabilistic description is impossible (because the complete information about a system is not available). 

In such a case we represent our partial knowledge about a system by the complex probabilistic amplitude. Such a complex probability amplitude $\psi(t, x)$ describes dynamics of context (of e.g., the financial market, or some branch of it, or a corporation). However, there arises the following natural question:

\medskip

{\it Can the financial market (or more generally, economics) be described as 
a quantum-like probabilistic system?}

\medskip

If the answer to this question is positive, then one should apply the mathematical
formalism of quantum probability theory to describe financial and economic processes. If 
it is negative, then one should be satisfied by the present approach which is based
on classical stochastic differential equations.

The only possibility to answer this question is to perform experimental statistical 
tests to verify non-classicality of financial or more general economic data. One of 
such tests is based on comparing classical and quantum formulas, (\ref {1}) and (\ref {2}). 
An experiment of such a type has been already done [4]. 
It confirmed the quantum-like behaviour of cognitive systems. We 
found the quantum-like interference of probabilities for answers to questions
which were asked to groups of students. We remind that in our quantum-like model the 
interference is not the interference of waves, but the interference of probabilities. 
The additional interference term in the formula (\ref {2}) is a consequence 
of dependence of the probabilities on contexts and not an exhibition of some 
mystical wave features of e.g., traders at the financial market. The 
quantum-like wave function represents contextual probabilistic features and nothing more.

In this note we describe another experimental statistical test which could be used to distinguish classical and quantum-like probabilistic behaviour in finances and economy.

The last decades quantum physics has been strongly engaged in research which was 
related to violations of Bell's inequality [17]. In the 
60th J. Bell proved some inequality which should be valid for correlations of classical quantities, 
but violated for some quantum correlations. There were performed experimental tests which demonstrated that Bell's inequality is really violated for experimental data.

Our idea is to use an analogue of Bell's test in finances and economics. 
However, it seems to be impossible to apply directly the physical methods to economics and finances.
The main problem was that in quantum physics Bell's inequality was tested for 
{\it pairs of entangled particle}. Roughly speaking these are particles-copies 
(at least with some approximation). Direct realization of  such a test with e.g. traders of 
the financial market is practically impossible. 

We choose another way. We move from the original Bell's inequality for correlations (or 
its analogue - Wigner's inequality - for joint probabilities) and prove an analogous 
inequality for conditional probabilities. Such a test based on conditional probabilities can 
easily be  performed with economic agents. If our Bell-like inequality 
for conditional probabilities should be violated, then such an experimental fact can be considered 
as a strong evidence in favor of quantumness of financial and economic processes.

\section{Bell-like inequalities}

Let ${\cal P}=(\Omega, {\cal F}, {\bf P})$ be a Kolmogorov probability space, 
[18]. For any pair of random variables $u(\omega), v(\omega),$ their covariation is defined by
$$
<u,v>= {\rm cov}(u,v)=\int_\Omega u(\omega) v(\omega) d{\bf P}(\omega).
$$

{\bf Theorem 2.1.} (Bell inequality for covariations) {\it Let $\xi_a, \xi_b, \xi_c= \pm1$ be random variables on {\cal P}. Then Bell's inequality
\begin{equation}
\label{3}
\vert <\xi_a, \xi_b> - <\xi_c, \xi_b> \vert \leq 1 - <\xi_a, \xi_c>
\end{equation}
 holds.}

{\bf Theorem 2.2.} (Wigner inequality) {\it Let $\xi_a, \xi_b, \xi_c= \pm1$ be arbitrary random variables on a Kolmogorov space {\cal P}. Then the following inequality holds: }
\begin{equation}
\label{4}
{\bf P}(\xi_a = +1, \xi_b= +1) + {\bf P}(\xi_b= -1, \xi_c= +1) \geq {\bf P}(\xi_a = +1, \xi_c= +1)
\end{equation}

The detailed proofs can be found e.g. in [3]. Bell's type inequalitites are 
applied in the following way. One prepares pairs of entangled particles $s = (s_1, s_2).$ 
There are observables related to the first andd seond particles. They are labeled by 
some paramaters taking values $\theta = a, b, c.$ Denote random variables corresponding
 to these observables\footnote{We would like to test the hypothesis that such classical 
random variables can be introduced -- so to test the 
possibility to apply the classical probabilistic model.} for the first particle by $\xi_a(\omega), \xi_b(\omega), \xi_c(\omega)$ and for the second by $\eta_a(\omega), \eta_b(\omega), \eta_c(\omega).$ Entanglement of particles implies the precise correlations (or anti-correlations - depending on statistics):
\begin{equation}
\label{5}
\xi_\theta(\omega)= \eta_\theta(\omega)
\end{equation}
for all values of the parameter $\theta$. Therefore we can place $\eta_\theta(\omega)$ to 
the second place (instead of $\xi_\theta(\omega)$) in the inequality (\ref{4}). Instead of nonphysical probabilities ${\bf P}(\xi_a = +1, \xi_b = +1), {\bf P}(\xi_b = -1, \xi_c = +1), {\bf P}(\xi_a = +1, \xi_c = +1),$ we obtain the physical ones ${\bf P}(\xi_a = +1, \eta_b = +1), {\bf P}(\xi_b = -1, \eta_c = +1), {\bf P}(\xi_a = +1, \eta_c = +1).$ The main idea (belonging to Einstein, Podolsky and Rosen) is that, although we are not able to make the second measurement on the same particle $s_1$ without to disturb it totally, nevertheless, we can measure $\xi_{\theta_1}$ on the first particle and $\eta_{\theta_2}$ on the second particle. We obtain the inequality which can be experimentally verified:
$$
{\bf P}(\xi_a = +1, \eta_b = +1)+ {\bf P}(\xi_b = -1, \eta_c = +1) \geq {\bf P}(\xi_a = +1, \eta_c = +1).
$$
The situation with financial and econoomic agents is very similar to the quantum one. 
They are also very sensitive to questions. Let now $\theta = a, b, c$ be three different questions.
 Of course, they should be really disturbing for agents. By giving the answer $\xi_a(\omega)$ 
the agent creates a new memory which will play an important role in answering the next
 question, e.g., $b$ or $c$. The main problem is that we are not able to prepare an ensemble 
of ``entangled agents", i.e., to satisfy the condition of precise correlations (\ref{5}).
 Therefore we shall change the strategy and operate with conditional probabilities. 
To find conditional probabilities, one need not operate with pairs of ``entangled agents" 
of the financial market. One can perform successive measurements (in the form of questions to agents).

As a simple consequence of Theorem 2.2, we obtain [3] the following mathematical result:

{\bf Theorem 2.3.} (Wigner inequality for conditional probabilities) {\it Let $\xi_a, \xi_b, \xi_c = \pm 1$ be symmetrically distributed random variables on {\cal P}. Then the following inequality holds true:}
\begin{equation}
\label{6}
{\bf P}(\xi_a = +1\vert \xi_b = +1) + {\bf P}(\xi_c = +1\vert \xi_b = -1) \geq {\bf P}(\xi_a = +1\vert \xi_c = +1).
\end{equation}

The latter equality can easily be  tested experimentally. We choose three 
``mutually disturbing questions,'' $a, b, c$ about finances or an 
economic situation. The answer "yes" is encoded by +1 and the answer "no" by -1. 
We prepare a homogeneous ensemble of people (with the 
same age, education, political and national background), say $S$. Then
 we divide it into three sub-ensembles, $S_1, S_2, S_3,$ of the same size (without  violating  
the homogeneous structure of the ensemble $S$).

By asking questions $b$ and $a$ in the first ensemble (the order plays a 
cruicial role) we shall find the frequency corresponding to the conditional probability ${\bf P}(\xi_a = +1\vert\xi_b = +1):$
$$
\nu(\xi_a = +1\vert \xi_b = +1)= \frac{n(\xi_a = +1\vert\xi_b = +1)}{N(\xi_b = +1)},$$
where $N(\xi_b = +1)$ is the number of agents in the ensemble $S_1$ who gave the answer "yes" to the question $b$ and $n(\xi_a = +1\vert\xi_b = +1)$ is the number of agents who gave the answer "yes" to the question $a$ among those who have already answered "yes" to the question $b$.

By asking questions $b$ and $c$ in the second ensemble (we repeat 
that the order of questions plays a cruicial role!) we shall find the frequency 
corresponding to the conditional probability ${\bf P}(\xi_c = +1\vert\xi_b = -1):$
$$
\nu(\xi_c = +1\vert \xi_b = -1)= \frac{n(\xi_c = +1\vert\xi_b = -1)}{N(\xi_b = -1)},
$$
where $N(\xi_b = -1)$ is the number of agents in the ensemble $S_2$ who gave the answer "no" to the question $b$ and $n(\xi_c = +1\vert\xi_b = -1)$ is the number of agents who gave the answer "yes" to the question $c$ among those who have already answered "no" to the question $b$.

By asking questions $c$ and $a$ in the third ensemble we shall find the frequency corresponding 
to the conditional probability ${\bf P}(\xi_a = +1\vert\xi_c = +1).$ Finally,
 we put those frequencies into the inequality (\ref{6}). If this inequality should 
be violated, we should obtain a strong argument supporting our the 
hypothesis about the quantum-like behaviour of financial or economic processes.

The experimental framework can even be  simplified. We can split the original ensemble $S$ 
into just two sub-ensembles $S_1$ and $S_2$. We ask the question $b$ to agents in the
 first ensemble. We create two new sub-ensembles with respect to the answers "yes" 
and "no" to this question, $S_{1, b= +1}$ and $S_{1, b= -1}$. Then we shall ask the
 $a$-question to agents in the ensemble $S_{1, b= +1}$ and the $c$-question to agents
 in the ensemble $S_{1, b=-1}.$ In this way we shall obtain the frequencies 
$\nu(\xi_a = +1\vert \xi_b = +1), \nu(\xi_c = +1\vert\xi_b = -1).$ 
By using the ensemble $S_2$ we obtain the frequency $\nu(\xi_a = +1\vert\xi_c = +1).$

The main preparation constraint for this experiment is that all quetions $a, b, c,$ should induce the symmetric probability distributions: ${\bf P}(\xi_a = +1) = {\bf P}(\xi_a = -1) = 1/2, {\bf P}(\xi_b = +1) = {\bf P}(\xi_b = -1) = 1/2, {\bf P}(\xi_c = +1) = {\bf P}(\xi_c = -1) = 1/2.$

\section{Efficient market hypothesis}
If the proposed test confirms our hypothesis about quantum-like probabilistic behaviour of the financial market, then it may have interesting consequences for foundations of financial mathematics. In the quantum-like approach the fundamental assumption of the modern financial mathematics, namely, the efficient market hypothesis [19], [20], [2] would be questioned.

The financial context (situation at the financial market, including expectations, prognoses, political situation, social opinion) $C_{\rm fin}(t)$ is represented by a complex probability amplitude, {\it financial wave function} $\psi(t, q),$ where $q$ is the vector of prices of shares (it has a huge dimension). The evolution of $\psi(t, q)$ is described by a {\it deterministic equation} - Schr\"odinger's equation. Hence, the evolution of the financial context $C_{\rm fin}(t)$ could be predicted - at least in principle. Of course, at the moment one could not even dream about the possibility to solve the problem analytically or even numerically. First of all there is no idea how ``financial Hamiltonian'' (quantum-like operator representing the `` energy of the financial market'') should be constructed. Another problem is the huge dimension of the problem. However, our quantum-like model provides the qualitative prediction that there might be developed financial technologies which induce permanently exploitable profit opportunities (in the opposition to the conventional model based on the efficient market hypothesis).

We emphasize that the possibility to create such quantum-like financial technologies does not imply lower complexity of our model comparing with the conventional one. The latter implies that financial processes can be represented by a special class of classical stochastic processes, martingales. For any such process we can construct a single Kolmogorov probability space for all realizations of this process (this is the essence of the famous Kolmogorov theorem [18]). In contrast to such a single space description, in our  quantum-like model one could not assume that the quantum-like process based on the evolution of the financial context could be embedded into a single Kolmogorov probability space.

In the classical financial mathematics there were performed fundamental investigations to find an 
adequate stochastic processes matching the real financial data: Brownian, geometric Brownian,
 general Levy processes. From the point of view of our quantum-like approach the 
problem cannot even be  formulated in such a way. There is not
any classical stochastic process which will match with the real financial data, 
because there is not a single Kolmogorov space describing the whole financial market.
The financial data can only be  represented as a quantum-like financial process.

Finally we remark that the efficient market hypothesis has been questioned by many authors, from
other points of view; see  e.g. [21].

\bigskip

{\bf References:}

[1] Bachelier L 1890 {\it Ann. Sc. lÉcole Normale Superiere} {\bf 111-17} 21.\\

[2] Shiryaev A N 1999 {\it Essentials of Stochastic Finance: Facts, Models, Theory} (Singapore: World Scientific Publishing Company).\\

[3] Khrennikov A Yu 1999 {\it Interpretations of Probability} 
(VSP Int. Sc. Publishers, Utrecht/Tokyo (second edition-- 2004)).\\

[4] Khrennikov A Yu 2004 {\it Information dynamics in cognitive, psychological, social, and anamalous phenomena} (Kluwer, Dordrecht).\\

[5] E. Haven, A discussion on embedding the Black-Scholes
option pricing model in a quantum physics setting, Physica A {\bf
304}, 507-524 (2002).

[6]  E. Haven, A Black-Sholes Schr\"odinger option price: `bit'versus `qubit',
Physica A {\bf 324}, 201-206 (2003).

[7]  E. Haven, An `$h$-Brownian motion' of stochastic
option pices, Physica A {\bf 344}, 151-155 (2003).

[8] Segal W and Segal I 1998 {\it Proc. Nat. Acad. Sc. USA} {\bf 95} 4072.\\

[9] Haven E 2006 Bohmian mechanics in a macroscopic quantum system {\it Foundations of Probability and Physics-3} vol. 810 (Melville, New York: AIP) p 330.\\

[10] Piotrowski E W Sladkowski J 2001 Quantum-like approach to financial risk: quantum anthropic principle {\it Preprint} quant-ph/0110046.\\

[11] Choustova O 2001 Pilot wave quantum model for the stock market, http://www.arxiv.org/abs/quant-ph/0109122.\\

[12] Choustova O 2004 Bohmian mechanics for financial processes, 
{\it J. Modern Optics} {\bf 51}, 1111.\\

[13] Choustova O 2006 {\it Physica A: Statistical Physics and its Applications}.
{\bf 374,} 304-314 (2006).

[14] von Neumann J 1955 {\it Mathematical foundations of quantum mechanics} (Princeton Univ. Press, Princeton, N.J.).\\

[15] d'Espagnat B 1995 {\it Veiled Reality. An analysis of present-day quantum mechanical concepts} (Addison-Wesley).\\

[16] Shimony A 1993 {\it Search for a naturalistic world view} (Cambridge Univ. Press).\\

[17] Bell J S 1987 {\it Speakable and unspeakable in quantum mechanics} (Cambridge Univ. Press, Cambridge).\\

[18] Kolmogoroff A N 1933 {\it Grundbegriffe der Wahrscheinlichkeitsrech} (Springer Verlag, Berlin); reprinted: 1956 {\it Foundations of the Probability Theory} (Chelsea Publ. Comp., New York).\\

[19] Samuelson P A 1965 {\it Inductrial Management Rev.} {\bf 6} 41.\\

[20] Fama E F 1970 {\it J. Finance} {\bf 25} 383.

[21] B. Mandelbrot, R. Hudson, {\it The (mis)behavior of markets. A fractal view of risk, ruin, 
and reward.} Basic Books Publ., 2004
\end{document}